# DolNet: A DIVISION OF LABOUR BASED DISTRIBUTED OBJECT ORIENTED SOFTWARE PROCESS MODEL

Sachin Lakra, Deepak Kumar Sharma

**Abstract**: Distributed Software Development today is in its childhood and not too widespread as a method of developing software in the global IT Industry. In this context, Petrinets are a mathematical model for describing distributed systems theoretically, whereas AttNets are one of their offshoots. But development of true distributed software is limited to network operating systems majorly. Software that runs on many machines with separate programs for each machine, are very few. This paper introduces and defines Distributed Object Oriented Software Engineering (DOOSE) as a new field in software engineering. The paper further gives a Distributed Object Oriented Software Process Model (DOOSPM), called the DolNet, which describes how work may be done by a software development organization while working on Distributed Object Oriented (DOO) Projects.

## 1. Introduction

Distributed computing involves breaking down a large problem into small components, each of which is solved by a number of computers communicating with each other through a network. Distributed computing software today is more at a research stage than at a stage that could be considered as commercial. Attempts have been made to develop various types of software. Software developed earlier, which is currently popular, includes Network Operating Systems such as Microsoft Windows, Novell NetWare and Linux. Large scale projects include distributed computing projects such as SETI@home and Einstein@home, which use the Internet to connect to millions of machines.

The project SETI@home involves scanning the skies using radio telescopes, located at a central location in the USA, to search for any possible radio signals generated by extra terrestrials. The amount of data collected is so huge that it is not possible for a single computer to process and analyze it. For this reason, the data is broken down into smaller chunks and uploaded on millions of machines connected through the Internet. Each machine processes the data and returns the results to the SETI@home servers, which in turn analyze it. Similarly, Einstein@home is a project which is trying to solve the problem of detecting gravity waves (predicted first by Albert Einstein) generated by pulsars, which are located at a great distance from the Earth.

As such, there is no software process model available for handling heterogeneity and scale during the development stages of a distributed software.

Section 1 introduces the paper. Section 2 of the paper defines the terms Software Engineering, Object Oriented Software Engineering, and Distributed Computing, and also defines the new field of Distributed Object Oriented Software Engineering. The paper further gives a Distributed Object Oriented Software Process Model (DOOSPM), called the DolNet, in Section 3, which describes how tasks may be completed by a software development organization while working on a Distributed Object Oriented (DOO) Project, focusing on solving the problems of heterogeneity of machines and operating systems, as well as that of scale. The generic approach taken in the DolNet involves the Division Of Labour (DOL) in a distributed manner among teams according to the types of *machines* and the types of *operating systems* that are going to run the DOO software. The work of a single team is independent of the *number* of machines, if the machines are of the same *type*. During coding, components are provided by a generic Object Oriented Component Library (OOCL). The OOCL may itself be a Distributed OOCL (DOOCL), if the project is very large. The paper then gives the advantages and disadvantages of the DolNet. Section 4 of the paper gives the conclusion.

## 2. Distributed Object Oriented Software Engineering (DOOSE)

Distributed Object Oriented Software Engineering (DOOSE), is a new component of the wide field of Software Engineering, which is being developed by the authors so as to bring some order to the field of distributed computing from a development methodology perspective. The definition for DOOSE is developed by building upon the existing related fields. Therefore, the authors first define these related fields.

## 2.1 Software Engineering

*Definition 1*: Software Engineering is defined as the establishment and use of sound engineering principles in order to economically obtain software that is reliable and works efficiently on real machines. [5]

*Definition 2*: Software Engineering is the application of a systematic, disciplined, quantifiable approach to the development, operation, and maintenance of software. [2]

## 2.2 Object Oriented Software Engineering (OOSE)

Object Oriented Programming (OOP) is based on the concept of components being assembled together to create a software. The concept is borrowed from the idea of integrated circuits or chips, resistors, capacitors, etc., being brought together on a printed circuit board to form a piece of hardware. The components in hardware can be picked "off-the-shelf", placed on a printed circuit board and then soldered onto it. Similar is the case with OOP where software components, i.e., predefined class definitions stored in a component library are "picked" and "placed" in a program. The "features" and "actions" of the software components, i.e., their properties and methods, respectively, are then used to implement the proposed software. This is the general approach with OOP. [4]

*Definition 3*: Object Oriented Software Engineering is the discipline of reusing existing components, or using, after creating, unavailable components to assemble a software by the application of object oriented principles.

*Definition 4*: An Object Oriented Component Library (OOCL) is a repository of predefined software components which can be "picked off-the-shelf" from the Component Library and "placed" in a program.

## 2.3 Distributed Computing

*Definition 5*: **Distributed computing** is a method of computer processing in which different parts of a program are run simultaneously on two or more computers that are communicating with each other over a network. [1]

Distributed computing is a type of **segmented** or parallel computing, but the latter term is most commonly used to refer to processing in which different parts of a program run simultaneously on two or more processors that are part of the same computer. While both types of processing require that a program be segmented—divided into sections that can run simultaneously, distributed computing also requires that the division of the program take into account the different environments on which the different sections of the program will be running. For example, two computers are likely to have different file systems and different hardware components.

**2.4 Distributed Object Oriented Software Engineering (DOOSE)**

*Definition 6*: Distributed Object Oriented Software Engineering is the application of object oriented principles to the development and implementation of distributed software, taking into account the heterogeneity of machines on which the distributed computing will take place.

*Definition 7*: A Distributed Object Oriented Component Library (DOOCL) is an Object Oriented Component Library, which makes software components available to teams, working on modules of a distributed object oriented project, in a distributed manner.

**3**. **The DolNet**

The DolNet is a Division of Labour (Dol) based Distributed Object Oriented Software Process Model (DOOSPM). The generic approach taken in the DolNet involves the division of labour in a distributed manner among teams according to the types of *machines* and the types of *operating systems* that are going to run the DOO software. The work of a single team is related to a single combination of an operating system and a machine. The work of a single team is independent of the *number* of machines, if the machines are of the same *type*. Such a division of labour allows rapid application development to be applied. During coding, object oriented components are provided by a Distributed Object Oriented Component Library (DOOCL).

**3.1 Notation**

The description of the DolNet approach as depicted in Figure 2 and Figure 3 uses the notation given in Figure 1.

An activity is represented by a rectangle, with the name of the activity written inside it. The undirected lines represent links among the activities. The directed arrows show the input of a resource to an activity, or the direction of flow from activity to activity, according to sequence. A hexagon represents a module. Resources are represented by a rectangle, with an alphabet or shape shown in a rectangular compartment on the left side of the rectangle. Alphabets in the compartments of a resource rectangle, include D for a

DOOCL, P for a program (as input to testing activity) and T for a test suite, whereas the shape of a human being represents a team. In the right hand side compartment, the ID and the name of the resource is written for identification.

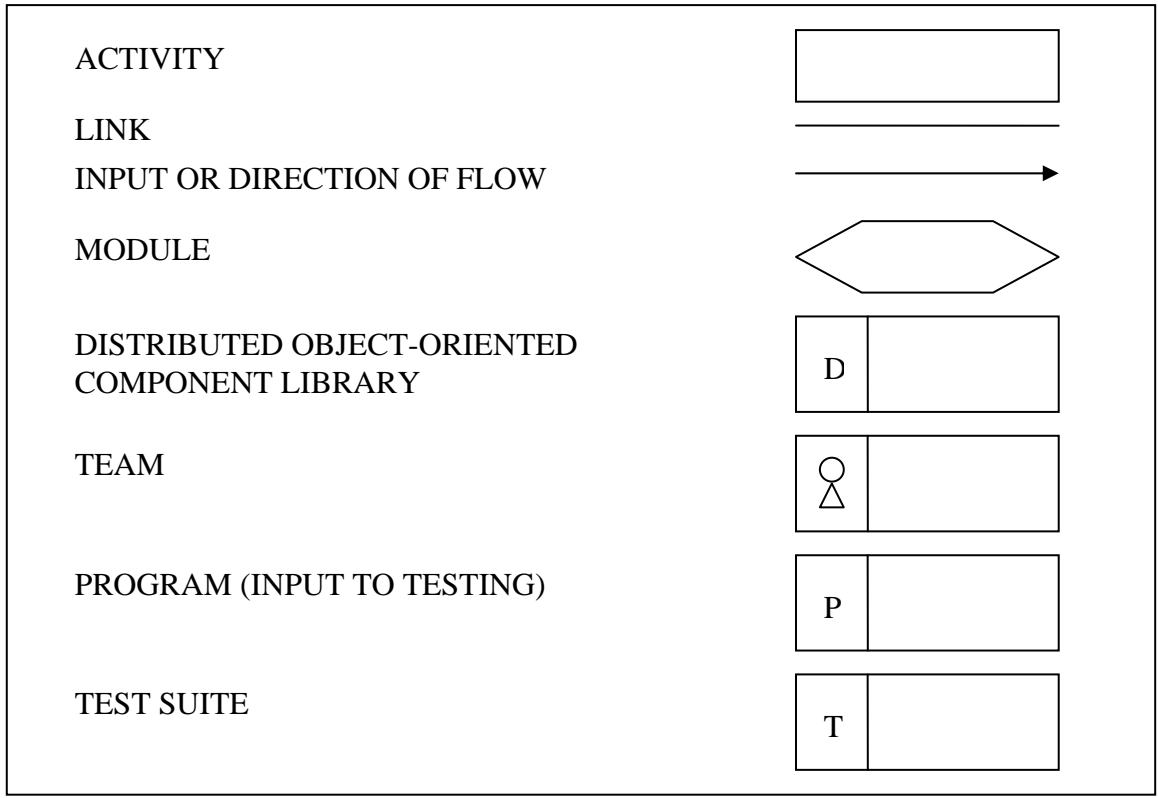

*Figure 1        : The notation used in the DolNet.*

## 3.2 The DolNet Distributed Object Oriented Software Process Model

The approach taken in the DolNet is a combination of the Waterfall Model and the Rapid Application Development (RAD) Model. The basic activities defined are based on the Waterfall Model as shown in Figure 2, and these activities are described below. The division of labour into teams is defined in the form of a number of Module Development Activities, each constituted by the coding and testing stages of the module for a single machine-OS combination or a geographically separate module. The activities are:

*Integrated Study:* The study of the DOO Project should be done in an integrated manner with the participation of at least one representative systems analyst from each team involved.

*Integrated Analysis:* The analysis should involve systems analysts and software engineers from each team working on the project with emphasis on understanding the system as an integrated whole.

*System Design:* The system design should take an overall view of the project and define the structure of the proposed system as a whole. This is the stage when work should be divided into modules based on the fact that the modules may be different either in their machine type and operating system combination or may be geographically dispersed forming a large project.

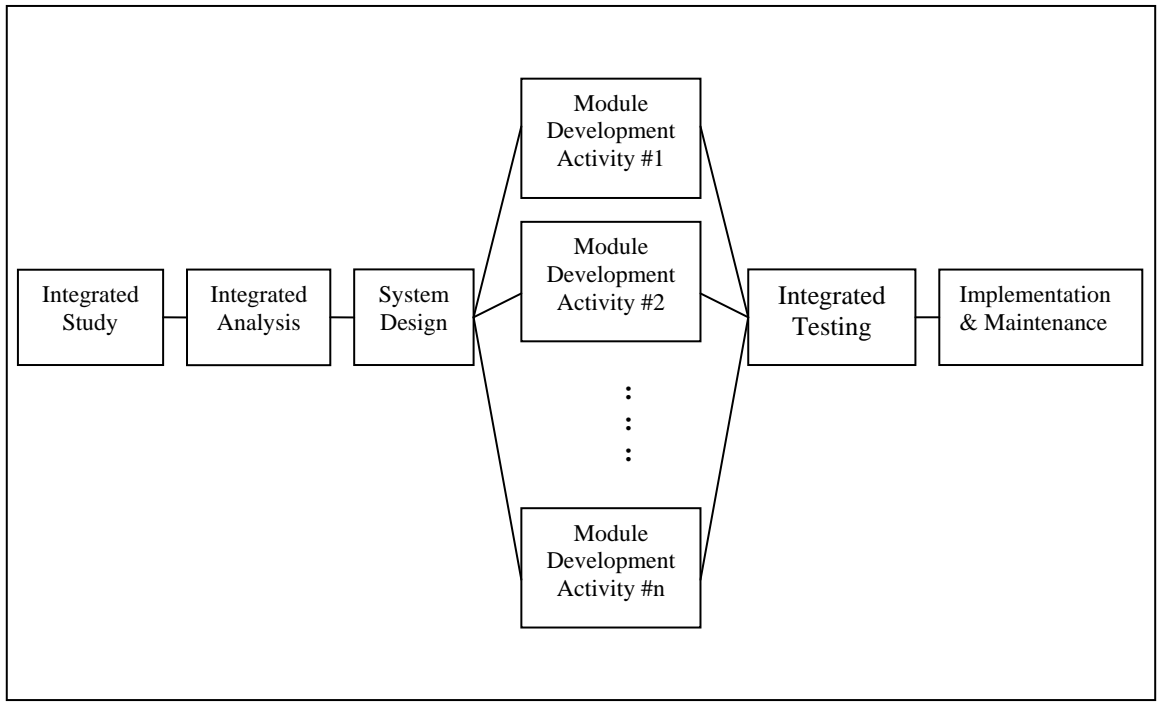

*Figure 2        :The DolNet DOOSPM.*

*Module Development Activity # n:* Each module is then assigned to a team, which uses a separate DOOCL for its tasks. The tasks included in this stage are coding of each of the programs in the module, and testing of each program using a test suite, followed by a Module Testing Activity for the complete module. The DOOCL is checked, during coding of a program, for components required by the program, which are extracted, if found, or are developed as new components, otherwise. New components are then used by the module and added to the DOOCL as well (Figure 3). The DOOCL updation activities are derived from the Component Assembly Model.

*Figure 3       : A Module Development Activity.*

*Integrated Testing:* The testing activity consists of applying various integration testing techniques on the integrated software developed as a result of the various Module Development Activities.

*Implementation & Maintenance:* This activity involves installation and implementation of the various modules on their respective machine-OS combinations after completion of integrated testing and acceptance by the client(s).

**3.3 Advantages and Disadvantages of DolNet**

The DolNet puts forward a new process model for a new field with the following advantages:

- It suggests a method for handling heterogeneity, which exists because of the availability of various types of machines, such as 16-bit or 32-bit or 64-bit machines, and because of various operating systems, such as MS Windows, Linux, etc.
- It also suggests a way of handling large scale Distributed Object Oriented Projects, by dividing the work of the project among a number of teams, whence the Rapid Application Development is applied to each of the individual modules.
- It applies the object oriented concepts to distributed computing in the form of a process model.

The disadvantage of the DolNet is that it is as yet untested, that is presently, it is only a theoretical concept.

**4. Conclusion**

The DolNet provides a novel strategy for the development of large scale distributed computing projects by applying the object oriented methodology to the type of projects known as Distributed Object Oriented Projects. A new language could possibly be developed to apply this process model. The authors will present a case study applying the DolNet in the near future.

**5. References**

[1] http://en.wikipedia.org/wiki/Distributed_computing

[2] "IEEE Standard Glossary of Software Engineering Terminology," IEEE standard 610.12-1990, 1990.

[3] Roger S. Pressman, "Software Engineering: A Practitioner's Approach", fifth edition, 2001.

[4] Sachin Lakra, Nand Kumar, Sugandha Hooda, Nitin Bhardwaj; "A Metric For The Activeness Of An Object-Oriented Component Library", Proceedings of World Conference 07, Las Vegas, Nevada, USA, April 2007

[5] Ubiquitously cited as a quote from F.L. Bauer at the original NATO Conference on Software Engineering, the usual citation being: (January 1969) "NATO Software Engineering Conference 1968" (pdf). edited by P. Naur and B. Randell; published January 1969 *Software Engineering: Report of a conference sponsored by the NATO Science Committee, Garmisch, Germany, 7-11 Oct. 1968*, Brussels: Scientific Affairs Division, NATO.